# All Optical Three Dimensional Spatio-Temporal Correlator for Automatic Event Recognition Using a Multiphoton Atomic System

MEHJABIN S. MONJUR,[1,*] MOHAMED F. FOUDA,[1] AND SELIM M. SHAHRIAR[1,2]

[1]Department of Electrical Engineering and Computer Science, Northwestern University, Evanston, IL 60208, USA
[2]Department of Physics and Astronomy, Northwestern University, Evanston, IL 60208, USA
*Corresponding author: mehjabin@u.northwestern.edu

________________________________________________________________

**Abstract**

We describe an automatic event recognition (AER) system based on a three-dimensional spatio-temporal correlator (STC) that combines the techniques of holographic correlation and photon echo based temporal pattern recognition. The STC is shift invariant in space and time. It can be used to recognize rapidly an event (e.g., a short video clip) that may be present in a large video file, and determine the temporal location of the event.  Using polar Mellin transform, it is possible to realize an STC that is also scale and rotation invariant spatially.  Numerical simulation results of such a system are presented using quantum mechanical equations of evolution. For this simulation we have used the model of an idealized, decay-free two level system of atoms with an inhomogeneous broadening that is larger than the inverse of the temporal resolution of the data stream.  We show how such a system can be realized by using a lambda-type three level system in atomic vapor, via adiabatic elimination of the intermediate state.  We have also developed analytically a three dimensional transfer function of the system, and shown that it agrees closely with the results obtained via explicit simulation of the atomic response.  The analytical transfer function can be used to determine the response of an STC very rapidly. In addition to the correlation signal, other nonlinear terms appear in the explicit numerical model. These terms are also verified by the analytical model. We describe how the AER can be operated in a manner such that the correlation signal remains unaffected by the additional nonlinear terms. We also show how such a practical STC can be realized using a combination of a porous-glass based Rb vapor cell, a holographic video disc, and a lithium niobate crystal.

***Keywords:*** Spatio-Temporal Correlator, Automatic Event Recognition System, Nonlinear Optical Signal Processing, Multiphoton Atomic System.

________________________________________________________________

## 1. Introduction

Automated target recognition (ATR) has been a very active field of research for several decades. Significant advances in ATR have been made using analog approaches employing holographic correlators, as well as computational approaches using dedicated digital signal processing (DSP) chips or software. However, these techniques are inadequate for the task of automatic event recognition (AER).  AER is defined as the task of identifying the occurrence of an event within a large video data base.  Consider, for example, a data base that contains video surveillance gathered by a camera-equipped drone or a satellite, which is monitoring a site for suspicious activities, such as a truck of a particular size entering or exiting a facility.  For data gathered over a few hours, this event may have occurred several times. The goal of an AER system is to determine if these events occurred, when they occurred, and how many times. In principle, this can be achieved by searching through each frame in the data base, and comparing them with reference images. This process is prohibitively time consuming, even with a very efficient optical image correlator, a software or a DSP based image recognition system. However, by employing the properties of atoms[1,2,3,4,5] it is possible to realize an AER that can recognize rapidly the occurrence of events, the number of events, and the occurrence times.  Separate aspects of the overall technology needed for the AER have also been demonstrated by us earlier[6,7,8,9,10]. In this paper, we describe quantitatively the design of an AER system using Rb vapor in nano-porous glass with paraffin coating and show the simulation results of such a system. The AER will be realized by the technique of spatio-temporal holographic correlation[3,4,11,12,13]. This combines the process of translation invariant spatial holographic correlation with the process of translation invariant temporal correlation. The resulting system is a Spatio-Temporal Correlator (STC).

Shift invariant in space and time, the STC can recognize rapidly an event that may be present in a video file, and determine the temporal location of the event. In general, modeling the STC requires determining the temporal dynamics of a large number of inhomogeneously broadened atoms, multiplexed with free-space wave propagation equations. Here, along with modeling the STC using the Schrodinger equation for the temporal evolution of atoms excited by optical fields, we show that the response of the STC can be determined by modeling the response of the interaction medium as a simple, three-dimensional, multiplicative transfer function in the spatio-temporal Fourier domain. We explain the physical origin of this model, and then establish the validity of this model by comparing its prediction with that determined via the quantum mechanical dynamics. We then show some examples of the response of the STC using both methods. We also address practical issues in realizing such an STC. First, we show that optically off-resonant excitation of three-level atoms in the Λ configuration is suitable for realizing the STC. Furthermore, we show how a combination of a paraffin coated, nano-porous Rb vapor cell, a holographic video disc, and a lithium niobate crystal can be used to realize the STC for practical use.

The rest of the paper is organized as follows. In Section 2, we summarize the concept of a translation invariant temporal correlator where we discuss the numerical and analytical modelling of such a system. Here, we derive the details of the analytic expressions for the one dimensional temporal transfer function which is the basis of describing the functionality of the STC. In Section 3, we describe the basic model of an automatic target recognition system based on a translation-invariant STC. In Section 4, we describe the explicit architecture of an STC and introduce the analytic model for it. In Section 5, we present results from numerical simulations of the stimulated photon echo, temporal correlator and the automatic event recognition system, by using the quantum mechanical equations of evolution for an ensemble of inhomogeneously broadened atoms and by using the analytical model. In Section 6, we describe how a practical automatic target recognition system can be realized. Concluding statements are made in Section 7.

## 2. Translation Invariant Temporal Correlator (TI-TC)

To illustrate the Translation Invariant Temporal Correlator (TI-TC), consider a medium that is inhomogeneously broadened, meaning that the atoms inside the medium have a range of resonant frequencies. In such a medium, one can record a temporal data sequence by using a uniform recording pulse separated in time [14,15,16,17]. We start by applying a write pulse. This is followed by the query data stream, with a certain time lag. The spectral-domain interference between the writing beam and the query data stream (which can be viewed also as a manifestation of the Alford and Gold Effect[4,11,18,19]) is encoded in the coherence produced in the atomic medium. When the reference data stream is applied to this system, a correlation peak is observed in a temporally shift invariant manner. The numerical and analytical modeling of such a system is discussed later in this section. Here, we have used an idealized, decay-free two level system of atoms with an inhomogeneous broadening that is larger than the inverse of the temporal resolution of the data stream. It can be shown that an off-resonant excitation in a three-level system is equivalent to this model[20,21].

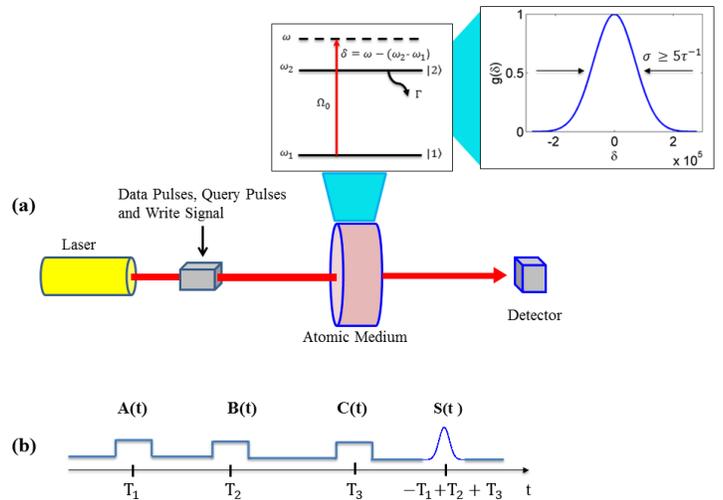

Fig 1: Schematic illustration of a temporal correlator using inhomogeneously broadened atomic media. After applying a short, uniform pulse of the writing beam, A(t) followed by the query, B(t) and reference data, C(t), a correlation peak, S(t) is observed in a temporal shift invariant manner.

## 2.1 Modelling of the Temporal Correlator using Two level Atoms (Numerical Model)

Consider an ensemble of two level atoms excited by a monochromatic field of frequency $\omega$. Here, $\hbar\omega_1$ and $\hbar\omega_2$ are the energies of levels $|1\rangle$ and $|2\rangle$ which are coupled by a laser field with a Rabi frequency of $\Omega_0$ and a detuning of $\delta$. The Schodinger equation under rotating wave approximation and transformation can be expressed as[20]:

$$i\hbar \, \partial |\Psi\rangle / \partial t = H |\Psi\rangle; \quad H/\hbar = -\delta |2\rangle\langle 2| + \Omega_0/2(|1\rangle\langle 2| + |2\rangle\langle 1|) \tag{1}$$

where $\delta \equiv \omega - (\omega_2 - \omega_1)$ and $|\Psi\rangle = c_1|1\rangle + c_2|2\rangle$. If we want to include the effect of decay due to spontaneous emission at the rate of $\Gamma$ from state $|2\rangle$, we must make use of the density matrix equation of motion[20]. However, the practical system we propose, as described in ref 21, is an effective two level system involving two metastable states. As such, we can set $\Gamma = 0$ in this model. This allows to us to make use of the amplitude equations (eqns. 1) to find the temporal evolution of each atom. The general solution of this equation can be expressed as follows:

$$\begin{bmatrix} c_1(t+\Delta t) \\ c_2(t+\Delta t) \end{bmatrix} = e^{\frac{i\delta t}{2}} \begin{bmatrix} \cos(\Omega'\Delta t/2) - i\delta/\Omega' \sin(\Omega'\Delta t/2) & -i\Omega_0/\Omega' \sin(\Omega'\Delta t/2) \\ -i\Omega_0/\Omega' \sin(\Omega'\Delta t/2) & \cos(\Omega'\Delta t/2) + i\delta/\Omega' \sin(\Omega'\Delta t/2) \end{bmatrix} \begin{bmatrix} c_1(t) \\ c_2(t) \end{bmatrix} \tag{2}$$

where, $\Omega' = \sqrt{\Omega_0^2 + \delta^2}$. To simulate the process of stimulated photo echo, for example, we start with an ensemble of two-level systems with a ground state $|1\rangle$ and an excited state $|2\rangle$, as depicted in the inset of figure 1(a). The spectral atomic distribution has a Gaussian profile with a width of $\sigma$ due to Doppler shift. The effective detuning seen by an atom moving with a velocity v in the direction of the laser beam is given by $\delta = \delta_o - kv$, where $\delta_o = \omega - \omega_o$ is the detuning of the laser for a stationary atom, and $\omega_o = \omega_2 - \omega_1$ is the resonance frequency of the atom. Note that, alternatively, this is equivalent to a laboratory frame picture in which the laser frequency is fixed at $\omega$, an atom with velocity v has a resonance frequency of $\omega_v = \omega_o + kv = (\omega_2 - \omega_1) + kv$, and the detuning experienced by this atom is $\delta = \omega - \omega_v = \omega - (\omega_o + kv) = \delta_o - kv$. For our simulations, we assume the laser to be resonant with the stationary atoms, so that $\omega = \omega_o$, and $\delta_o = 0$.

We suppose that, initially, all N atoms are prepared in the state $|1\rangle$. We first determine the quantum state of a band of atoms, with velocity v, after it has interacted with several laser pulses in sequence. This state is then used to determine the amplitude and phase of the induced dipole moment [proportional to $\rho_{12} \equiv c_1(t)c_2^*(t)e^{i[\delta t+(\omega_2-\omega_1)t]}$] oscillating at the frequency of $\omega_v = \omega_o + kv$. We then calculate the response of all the atoms with different velocities and add them together, weighted by the Gaussian distribution as a function of velocity. The resulting net dipole moment induced by the photon echo is given by: $P(t_4) = \int_{-\infty}^{\infty} \rho_{12}(t_4,\delta) g(\delta) d\delta$; where $g(\delta)$ is the Gaussian spectral distribution. The electric field of the resulting optical pulse is proportional to this net dipole moment. To ensure high-fidelity recording of the pulses in the atomic medium, we constrain the duration of the shortest light pulse to be at least five times longer than the inverse of $\sigma$. The temporal correlation process is illustrated schematically in figure 1. Figure 1(a) shows the simple architecture of the temporal correlator and figure 1(b) shows the sequence of pulses applied in an inhomogeneously broadened atomic medium. The simulation results using the quantum mechanical amplitude equations are described in details in section 5.

## 2.2 Derivation of the Transfer Function for the Temporal Correlation (Analytical Model)

To illustrate the temporal correlation process mathematically, we derive here the transfer function of the temporal correlator. In a later section, we show that this model is in close agreement with the response determined by solving the equations of motion of the atomic medium explicitly.

Let us consider the case where three temporal signals denoted as A(t), B(t) and C(t), are encoded on a laser beam with a modulator, as shown in figure 1(b). These functions represent the complex envelope of the electric field amplitude, with a central frequency of $\omega_L$. Explicitly, we can write:

$$E_Q(t) = Q(t)\exp(i(\omega_L t - kz)) + cc = |Q(t)|\exp(i\phi_Q); \quad (Q = A, B, C) \tag{3}$$

After applying the rotating wave approximation and the rotating wave transformation[20] (which is augmented to transform out the common phase factor *kz* as well), we find that the effective Hamiltonian for each of these fields can be expressed as:

$$H_Q(t)/\hbar = \omega|2\rangle\langle 2| + \Omega_Q(t)/2|1\rangle\langle 2| + \Omega_Q^*(t)/2|2\rangle\langle 1|; \quad (Q = A, B, C) \tag{4}$$

where the complex and time dependent Rabi frequency for each field is given by: $\Omega_Q(t) = \mu Q(t) = |\Omega_Q(t)|\exp(i\phi_Q), (Q = A, B, C)$; with $\mu$ being the dipole moment of the two level system, and the detuning of the center frequency of the laser ($\omega_L$) from the resonance frequency of the atom ($\omega_{Atom}$) is defined as $\omega \equiv \omega_{Atom} - \omega_L$.

As shown in figure 1(b), the three temporal signals have finite durations in time, and are separated from one another. Specifically, we assume that the three signals, A, B and C, arrive at the atomic medium at t=$T_1$, $T_2$ and $T_3$, respectively. Therefore, the Rabi frequencies seen by the atomic medium can be expressed as $\Omega_q(t) = \mu q(t) = |\Omega_q(t)|\exp(i\phi_Q); (Q = A, B, C; q = a, b, c)$, where $a(t) = A(t - T_1); b(t) = B(t - T_2); c(t) = C(t - T_3)$. Before proceeding further, we define explicitly the time domain Fourier Transform, $\tilde{g}(\omega)$ of a function $g(t)$ as follows:

$$g(t) = \frac{1}{\sqrt{2\pi}}\int_{-\infty}^{\infty}\tilde{g}(\omega)\exp(-i\omega t)d\omega; \quad \tilde{g}(\omega) = \frac{1}{\sqrt{2\pi}}\int_{-\infty}^{\infty}g(t)\exp(i\omega t)dt \tag{5}$$

From this definition, it then follows immediately that:

$$\tilde{\Omega}_a(\omega) = \mu\tilde{A}(\omega)\exp(i\omega T_1); \quad \tilde{\Omega}_b(\omega) = \mu\tilde{B}(\omega)\exp(i\omega T_2); \quad \tilde{\Omega}_c(\omega) = \mu\tilde{C}(\omega)\exp(i\omega T_3) \tag{6}$$

In the time domain, the atoms see the pulses at different times. However, the equivalent picture in the frequency domain is that the atoms see the Fourier components of all the pulses *simultaneously*, during the time window within which all three pulses are present. Thus, for $t \geq T_3$, the response of the atomic medium can be computed by assuming that it has interacted with all the fields simultaneously. To evaluate this response, we denote first as $N(\omega)$ the distribution of the atomic frequency detunings (i.e., the inhomogeneous broadening). Thus, the quantity $N(\omega)d\omega$ represents the number of atoms which have detunings ranging from $\omega - d\omega/2$ to $\omega + d\omega/2$, representing a spectral band of width $d\omega$. In the spectral domain view, a good approximation to make is that the atoms interact only with those components of the field that are resonant with the atoms, within a small band, justified by the fact that the spectral component within a vanishingly small band, for short enough pulses, is very small. Thus, the Schrodinger Equation for the amplitude of *this band of atoms*, in the rotating wave frame, is given by:

$$\frac{\partial}{\partial t}\begin{bmatrix} C_1(\omega) \\ C_2(\omega) \end{bmatrix} = -i\begin{bmatrix} 0 & \tilde{\Omega}(\omega)/2 \\ \tilde{\Omega}^*(\omega)/2 & 0 \end{bmatrix}\begin{bmatrix} C_1(\omega) \\ C_2(\omega) \end{bmatrix} \tag{7}$$

where the net, complex Rabi frequency within this band is given by
$\tilde{\Omega}(\omega) = \tilde{\Omega}_a(\omega) + \tilde{\Omega}_b(\omega) + \tilde{\Omega}_c(\omega) = \mu(\tilde{a}(\omega) + \tilde{b}(\omega) + \tilde{c}(\omega)) \equiv |\tilde{\Omega}(\omega)|\exp(i\phi(\omega))$

Assuming that all the atoms are in the ground state before the first pulse is applied, the solution for this equation, physically valid for $t \geq T_3$, is given by: $C_1(\omega) = Cos(|\tilde{\Omega}(\omega)|t/2); C_2(\omega) = -iSin(|\tilde{\Omega}(\omega)|t/2)\exp(i\phi(\omega))$. The amplitude of the electromagnetic field produced by the atoms *in this band* is proportional to the induced dipole moment, which in turn is proportional to the induced coherence, given by:

$$\rho_{12}(\omega,t) = C_1 C_2^* \exp(-i\omega_{atom}t) = C_1 C_2^* \exp(-i\omega_L t - i\omega t) \qquad (8)$$
$$= (i/2)\exp(-i\omega_L t)Sin[|\tilde{\Omega}(\omega)|t]\exp(-i\omega t) \times \exp(i\phi(\omega))$$

As we argued above, the component of the Rabi frequency within a very small band is very small, so that we can make use of the approximation that $Sin(\theta) \approx \theta - \theta^3/6$. Noting that the interaction occurs for a time window of duration $T \approx T_3 - T_1$, we can thus write that

$$\rho_{12}(\omega,t) \approx (i/2)\exp(-i\omega_L t) \times [\tilde{\Omega}(\omega)T - |\tilde{\Omega}(\omega)|^2 \tilde{\Omega}(\omega)T^3/6]\exp(-i\omega t) \qquad (9)$$

The signal (i.e., the electric field) produced by all the atoms can be expressed as:

$$\Sigma(t) = \alpha \exp(-i\omega_L t)\int_{-\infty}^{\infty} d\omega N(\omega)[\tilde{\Omega}(\omega)T - |\tilde{\Omega}(\omega)|^2 \tilde{\Omega}(\omega)T^3/6]\exp(-i\omega t) \qquad (10)$$

where the proportionality constant, $\alpha$, depends on the dipole moment of the two level system and the density of the atomic medium. For extracting the essential result, we assume that the width of the atomic spectral distribution is very large compared to that of $\tilde{\Omega}(\omega)$, so that $N(\omega)$ can be replaced by a constant, N. Furthermore, we define $\Sigma'(t) = \Sigma(t)\exp(i\omega_L t)$ as the envelope of the signal centered at the laser frequency, and $\beta = -\alpha N$, so that we can write:

$$\Sigma'(t) = \beta \int_{-\infty}^{\infty} d\omega [|\tilde{\Omega}(\omega)|^2 \tilde{\Omega}(\omega)T^3/6 - \tilde{\Omega}(\omega)T] \times \exp(-i\omega t) \qquad (11)$$

Note that the time dependent value of the off diagonal density matrix element $\rho_{12}(t)$, integrated over all atoms, is simply proportional to this signal: $\rho_{12}(t) = \xi\Sigma'(t)$, where $\xi$ is a proportionality constant. This, of course, is proportional to the density matrix element in the rotating wave frame:

$$\tilde{\rho}_{12}(t) \equiv \rho_{12}e^{i\omega_L t} = \xi\Sigma'(t)e^{i\omega_L t} \qquad (12)$$

The linear terms in eqn, 11 represent the so-called free-induction decay which occurs immediately after each pulse leaves the atomic medium, as can be shown easily, and do not contribute to the correlation signal. Since the net Rabi frequency has three components, corresponding to the three pulses, there will be a total of twenty seven components corresponding to the non-linear term. However, some of these terms are identical to one another except for numerical coefficients, leading to eighteen distinct terms. They are tabulated in table 1, where the coefficient in front of each term indicates the number of times it occurs. These terms can first be divided into two categories: causal and acausal. The acausal terms occur at a time that is earlier than the time of application of at least one of the three constituent input signals. These appear due to the fact that we have used Fourier Transforms instead of Laplace Transforms in our analysis. It is to be understood that these terms are unphysical. However, using Fourier Transform in the analysis of the temporal correlator makes it convenient to mutiplex with the spatial correlator for use in 3D STC, as illustrated in section 4.

The causal category can be broken up into two groups: those appearing at $t \leq T_3$ and those appearing after $t > T_3$. For reference, we thus have three different groups of signals, designated as follows: Group A: Causal and appearing before or at $T_3$; Group B: Acausal; Group C: Causal and appearing after $T_3$. This grouping is indicated in the lower right corner of Table 1. In grouping these terms, we have assumed that $(T_2 - T_1) < (T_3 - T_2)$. Under the assumptions made here in deriving these results, the only physically meaningful terms are those in Group C, since we are calculating the response of the atoms to the combined field of all three pulses. For the STC, the only relevant terms are also those in group C. These correspond to Inverse Fourier Transform of $\tilde{a}^*\tilde{b}\tilde{c}, \tilde{b}^*\tilde{c}^2, \tilde{a}^*\tilde{c}^2$, appearing at time $t = -T_1 + T_2 + T_3$, $t = -T_2 + 2T_3$ and $t = -T_1 + 2T_3$. The term that corresponds to the desired correlation signal is $\tilde{a}^*\tilde{b}\tilde{c}$. In section 5, we discuss in detail how to choose various parameters in a way that ensures no overlap between this correlation signal and the signal corresponding to the other two nonlinear terms in group C.

**Table 1: List of nonlinear terms from third order expansion**

|  | | Terms Appearing at $t \leq T_3$ | | Terms Appearing at $t > T_3$ | |
|---|---|---|---|---|---|
|  | | Nonlinear Terms | Temporal Position, t | Nonlinear Terms | Temporal Position, t |
| Causal | | $\tilde{a}^2\tilde{a}^*$ | $T_1$ | $2\tilde{a}^*\tilde{b}\tilde{c}$ | $-T_1+T_2+T_3$ |
| | | $\tilde{b}^2\tilde{b}^*$ | $T_2$ | $\tilde{a}^*\tilde{c}^2$ | $-T_1+2T_3$ |
| | | $\tilde{c}^2\tilde{c}^*$ | $T_3$ | $\tilde{b}^*\tilde{c}^2$ | $-T_2+2T_3$ |
| | | $2\tilde{a}\tilde{a}^*\tilde{b}$ | $T_2$ | | |
| | | $2\tilde{b}\tilde{b}^*\tilde{c}$ | $T_3$ | | |
| | | $2\tilde{a}\tilde{a}^*\tilde{c}$ | $T_3$ | | |
| | | $\tilde{a}^*\tilde{b}^2$ | $-T_1+2T_2$ | | |
| Acausal | | $\tilde{a}^2\tilde{b}^*$ | $(2T_1-T_2) < T_1$ | | |
| | | $\tilde{a}^2\tilde{c}^*$ | $(2T_1-T_3) < T_1$ | A: Causal & Appearing at $t \leq T_3$ | |
| | | $\tilde{b}^2\tilde{c}^*$ | $(2T_2-T_3) < T_2$ | | |
| | | $2\tilde{a}\tilde{b}\tilde{c}^*$ | $(T_1+T_2-T_3) < T_1$ | B: Acausal | |
| | | $2\tilde{a}\tilde{b}\tilde{b}^*$ | $T_1$ | | |
| | | $2\tilde{b}\tilde{c}\tilde{c}^*$ | $T_2$ | C: Causal & Appearing at $t > T_3$ | |
| | | $2\tilde{a}\tilde{c}\tilde{c}^*$ | $T_1$ | | |
| | | $2\tilde{a}\tilde{b}^*\tilde{c}$ | $T_1-T_2+T_3 < T_3$ | | |

We now consider explicitly the signal produced by the term $\tilde{a}^*\tilde{b}\tilde{c}$. It can be expressed as: $\Sigma'_C(t) = [\beta\mu^3/6]\int_{-\infty}^{\infty} d\omega[\tilde{a}^*(\omega)\tilde{b}(\omega)\tilde{c}(\omega)]\exp(-i\omega t)$. For simplicity, we now define the normalized signal as $\sigma(t) \equiv \Sigma'_C \sqrt{2\pi} * 6/[\beta\mu^3]$, so that we can write:

$$\sigma(t) = 1/\sqrt{2\pi} \int d\omega[\tilde{a}(\omega)\tilde{b}(\omega)\tilde{c}(\omega)]\exp(-i\omega t) \tag{13}$$

It then follows that the FT of the normalized correlation signal is:

$$\tilde{\sigma}(\omega) = \tilde{a}^*(\omega)\tilde{b}(\omega)\tilde{c}(\omega) = \tilde{A}^*(\omega)\tilde{B}(\omega)\tilde{C}(\omega)\exp(j\omega(T_3+T_2-T_1)) \tag{14}$$

If we define $\tilde{S}(\omega) = \tilde{A}^*(\omega)\tilde{B}(\omega)\tilde{C}(\omega)$ then then it follows that S(t) is the cross-correlation between A(t) and the convolution of B(t) and C(t). Since A(t) is essentially a delta function in time, S(t) is effectively the convolution of B(t) and C(t). Explicitly, if we consider $A(t) = A_0\delta(t)$, we get $S(t) = A_o\int_{-\infty}^{\infty} B(t')C(t-t')dt'$.

If we take into account the finite temporal width of A(t), this signal S(t) will be broadened by this added width. Finally, we note that $\sigma(t) = S(t-(T_3+T_2-T_1))$, which means that this correlation signal occurs at $t = T_3+(T_2-T_1)$, as already noted above. These results have been compared with actual numerical simulation in section 5, with close agreement.

## 3. Automatic Event Recognition (AER) via Translation-Invariant Spatio-Temporal Correlator

Over recent years, our group has been investigating the feasibility of realizing a high-speed automatic target recognition system using the Translation Invariant Spatial Holographic Correlation technique (TI-SHC)[22,23,24,25,26]. Using images pre-processed via so-called polar Mellin transforms, it is also possible to make the system invariant with respect to scale and rotation[27,28,29], while retaining the shift invariance. The natural extension to searching for images in spatial domain and searching for signals in time is to search for an image that is changing in time, which is simply a video clip corresponding to an event. Consider a video signal, either from a live camera feed or from a DVD player, for example. The video signal is an image that is changing in time, i.e. it has both spatial and temporal properties. An AER system can recognize a short clip within that video feed, by using a Translation Invariant Spatio-Temporal Correlator (TI-STC) which is a combination of the TI-SHC and the TI-TC.

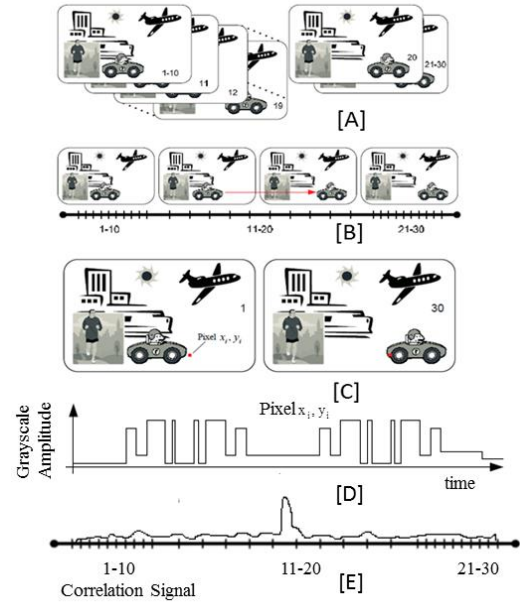

Fig 2: Schematic illustration of the automatic event recognition process. See text for details.

The AER process is illustrated schematically in figure 2. In figure 2(a), we show a series of 30 consecutive frames in a video. Frames 1-10 are of a static, unchanging scene. In frame 11 a car begins to drive across the scene, and stops in frame 20. Frames 21-30 are again static. This is illustrated in figure 2(b). If we consider a single pixel from this video signal, as indicated in figure 2(c), the resulting pattern as the car moves from one place in frame 11 to another in frame 20 is shown in figure 2(d). This signal is akin to the signals we used in the temporal signal correlation example in section 2. Each pixel in the video will have a corresponding bit stream, which could all be recognized separately in a temporal correlator. However, by combining the spatial correlator, we can now recognize a group of pixels that form an image, as they change in time.

As an example, consider the case where the event to be recognized is the car driving from one place to another. In the AER system, the whole video will be the database, and the ten clips corresponding to the movement of the car will be the query event. The output correlation signal in the AER system will contain a correlation peak corresponding to the time between frames 15 and 16, where the middle of the query event occurs, as illustrated in figure 2(e). Because the system is translation invariant spatially, we can find the car driving across the scene no matter where it occurs within the frame. Furthermore, because the system is translation invariant in time as well, no a priori knowledge regarding the start time of the query event within the database video is necessary. Furthermore, the time at which the correlation peak will appear can be used to infer the location of the query event within the database video. It should also be noted that if the same event occurs N times within the database video, N different correlation peaks will be observed. Finally, if the polar Mellin transform [26,27,28,29] is used to pre-process each of the frames in the database video as well as the query video, it would be possible to recognize the event even when the images in the query clip are scaled and rotated with respect to the database video.

## 4. Architecture of the Spatio-Temporal Correlator (STC)

The experimental configuration for realizing the AER system is illustrated schematically in figure 3. The architecture for the AER system is similar to that of a conventional spatial holographic correlator except that the write pulse is replaced by a plane wave of certain duration in space and time. The recording medium is replaced by the inhomogeneously broadened atomic medium (AM). The laser beam is directed to the reflection mode SLM with a polarizing beam splitter. The SLM reflects a pattern of light that is orthogonally polarized so that it passes through the same beam splitter. The pattern produced by the SLM is controlled by signals applied to it. The lens after the SLM produces the two-dimensional spatial FT of the SLM pattern in the plane of the atomic medium. Similarly, the second lens produces the two dimensional spatial FT of the field, generated by the atomic medium, in the plane of the detector array. The recording pulse is a highly localized spot in the plane of the SLM. For example, if it is located at $\{x = 0; y = y_o\}$, (with $y_o \ll L$), then, in the plane of the atomic medium (AM), it will appear essentially as a plane wave, moving in the y-z plane, and making an angle of $\sin^{-1}(a/L)$ with the z-direction. Temporally, the duration of the recording pulse is chosen to be much shorter than the temporal separation between consecutive frames in the query event or in the database video. This ensures that its spectrum will be much wider than that of the query clip and the database video. The timing sequence of the pulses is as follows. The recording pulse is applied first, and the time of the arrival of the center of this pulse at the AM is defined as $T_1$. Following a delay, the frames corresponding to the query event are sent to the SLM. The time of arrival of the center of this sequence at the AM is defined as $T_2$. As noted above, the first lens produces a two-dimensional spatial FT of each frame in the plane of the AM. Thus, the interference of the recording pulse and the query frames are stored in the AM, in the form of spatio-spectral gratings.

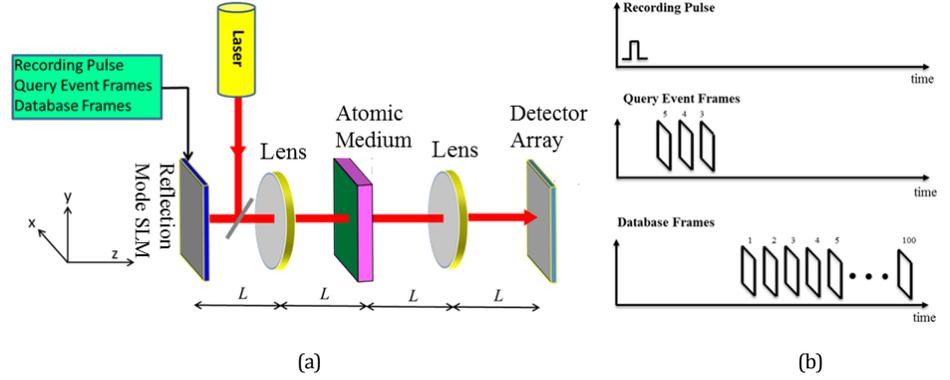

Fig 3: (a) Schematic illustration of the experimental configuration for an Automatic Event Recognition system employing spatio-temporal correlator. The focal length of each lens is L. (b) The sequence at which the recording pulse, reference and query film are applied to the atomic media.

After another delay, the database video frames are sent to the SLM, and the time of arrival of the center of this sequence at the AM is defined as $T_3$. Again, the first lens produces a two-dimensional spatial FT of each frame in the plane of the AM. These frames effectively diffract from the spatio-spectral gratings generated by the interference between the recording pulse and the query frames. The resulting signal from the AM passes through the second lens and the detector array records the output signals as functions of time. The output signal, integrated spatially over the detector array, contains a correlation peak at a time corresponding to the position where the matching pattern occurs in the database video. This is, of course, the desired functionality of the AER.

To illustrate the spatio-temporal correlation process mathematically, let us assume that the envelope of the electric field amplitude of the three sets of signals (recording beam, query clip, and database clip) and the resulting correlation signal can be expressed as $A(x,y,t)$, $B(x,y,t)$, $C(x,y,t)$, and $S(x_s, y_s, t)$, respectively, where $\{x, y\}$ are the transverse coordinates in the plane of the SLM, and $\{x_s, y_s\}$ are the transverse coordinates in the plane of the detector array. We assume further that, in the reference frame of the atomic medium, these pulses are centered at times $T_1$, $T_2$, $T_3$, and $T_4$ respectively. Denoting the temporal frequency as $\omega_T$, and the spatial frequencies as $k_x$ and $k_y$, it then follows that the three-dimensional FT's of these envelopes are given, respectively, by $\tilde{a}(k_x,k_y,\omega_T) = \tilde{A}(k_x,k_y,\omega_T)\exp(j\omega_T T_1)$, $\tilde{b}(k_x,k_y,\omega_T) = \tilde{B}(k_x,k_y,\omega_T)\exp(j\omega_T T_2)$, $\tilde{c}(k_x,k_y,\omega_T) = \tilde{C}(k_x,k_y,\omega_T)\exp(j\omega_T T_3)$ and $\tilde{\sigma}(k_x,k_y,\omega_T) = \tilde{S}(k_x,k_y,\omega_T)\exp(j\omega_T T_4)$. As derived in the Appendix A, using slightly different but equivalent notation, these FT's obey the following relation (ignoring an overall proportionality constant):

$$\tilde{\sigma}(k_x, k_y, \omega_T) = \tilde{a}^*(k_x, k_y, \omega_T)\tilde{b}(k_x, k_y, \omega_T)\tilde{c}(k_x, k_y, \omega_T) \tag{15}$$

$$\tilde{S}(k_x, k_y, \omega_T) = \tilde{A}^*(k_x, k_y, \omega_T)\tilde{B}(k_x, k_y, \omega_T)\tilde{C}(k_x, k_y, \omega_T) \tag{16}$$

It then immediately follows that the correlation signal appears at time $T_4 = T_3 + T_2 - T_1$. Since the write pulse is very short and uniform, it is effectively a delta function in time and position, and its three-dimensional FT is essentially uniform over the spatio-temporal spectral extent of the FT's of the data and query pulses, with a peak value of $A_o$. Under this condition, we have $\tilde{S}(k_x, k_y, \omega_T) \approx A_o \tilde{B}(k_x, k_y, \omega_T)\tilde{C}(k_x, k_y, \omega_T)$. Thus, the envelope of the signal is proportional to the three-dimensional, spatio-temporal convolution of the query frames and the database frames. Explicitly, we can write:

$$S(x_s, y_s, t) = A_o \int_{-\infty}^{\infty} dt' \int_{-\infty}^{\infty} dx' \int_{-\infty}^{\infty} dy' B(x', y', t') C(x_s - x', y_s - y', t - t') \tag{17}$$

As noted above, the correlation signal produced in the detector plane can also be calculated explicitly by solving the equation of motion for the atoms explicitly. For a large database, such a computation is exceedingly time consuming, since the atomic medium is inhomogeneously broadened. However, the analytical model presented here makes the computation much faster.

## 5. Numerical Simulations of the Stimulated Photon Echo, Temporal Correlator and Automatic Event Recognition System

In section 2 through 4, we have introduced the numerical and the analytical model and claimed that they are essentially equivalent. In this section, we verify this claim by comparing the simulation results of both models. For simplicity, first we show the simulation of a stimulated photon echo (SPE) process which is the simplest version of the temporal correlator.

Figure 4(a) shows the sequence of pulses associated with the SPE. A short pulse a(t) is applied as the writing beam at time $T_1$, followed by the query pulse b(t) at time $T_2$. At time $T_3$, the reference pulse is applied to this memory. An echo pulse is observed in a temporally shift invariant manner at time $t = -T_1 + T_2 + T_3$. If the SPE process is viewed as a temporal correlator, then the echo pulse represents the correlation peak. The photon echo process described above is simulated using the quantum mechanical amplitude equations (details in section 2.1), and the result is shown in figure 4(b). Here, we have used an idealized, decay-free two level system of atoms with an inhomogeneous broadening that is atleast five times larger than the inverse of the temporal width of each of the input pulse (all of which have the same width). The atoms are initially prepared in the ground state, so that we have $c_1(t_0 = 0) = 1$ and $c_2(t_0 = 0) = 0$. Here, the first $\pi/10$ pulse is followed, at time $T_1$, by two other $\pi/10$ pulses appearing at $T_2$ and $T_3$ respectively. The echo appears at $t = T_2 + (T_3 - T_1)$. The simulation of the atomic model has been performed in a super computer for faster calculation. In addition to the correlation term other nonlinear terms appear in the simulation which have been discussed in detail in section 2.2. In the analysis of section 2.2, we have shown that out of twenty seven nonlinear terms only three terms are relevant which correspond to the Inverse Fourier Transform of $\tilde{a}^*\tilde{b}\tilde{c}, \tilde{b}^*\tilde{c}^2, \tilde{a}^*\tilde{c}^2$ and they are categorized in group C (in Table 1). In the simulation of the temporal correlator using the numerical model in figure 4(b), we see that the above mentioned terms are appearing at time $t = -T_1 + T_2 + T_3$, $t = T_2 + 2T_3$ and $t = -T_1 + 2T_3$. Here, the desired correlation signal, which is denoted as s(t), corresponds to the term $\tilde{a}^*\tilde{b}\tilde{c}$ and it appears at time $t = -T_1 + T_2 + T_3$ as expected. It should be noted that there is a signal that appears at time $t = -T_1 + 2T_2$. This can be understood as the conventional photon echo signal. Since it appears before $t = T_3$, it is of no interest for the temporal correlation process.

To simulate the SPE using the analytical model, it is necessary to modify the transfer function of eqn. 14 by adding additional non-linear terms in Table 1. As we have discussed in detail in section 2.2, the only terms in Table 1 that are physically meaningful are those appearing in group C, which occur at $t > T_3$. Thus, the modified version of the transfer function can be expressed as:

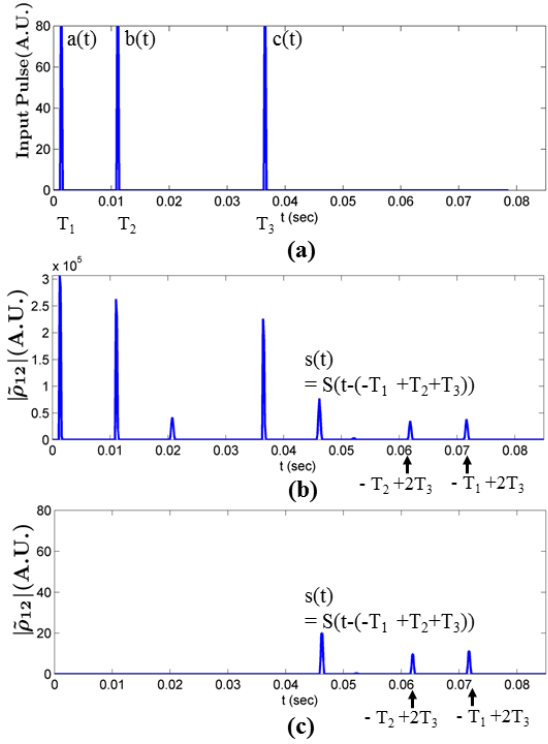
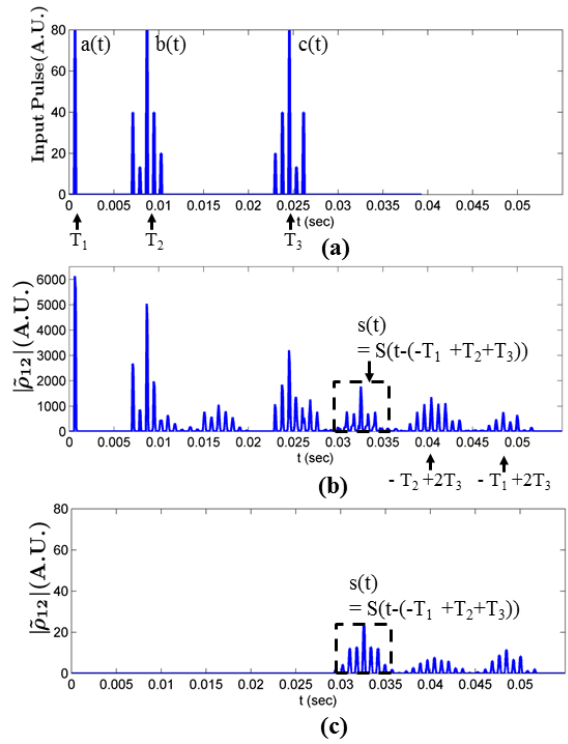

Fig 4: (a) Pulse sequence associated to the three pulse photon echo in two-level atoms. (b) Simulation results of photon echo using the numerical model which employs quantum mechanical amplitude equation and (c) Simulation results using the analytical model. Both model yields the same result with faster simulation time for the analytical model. [A.U. = Arbitrary Unit]

Fig 5: (a) Pulse train associated to the temporal correlator. (b) Simulation results of a temporal correlator using the numerical model which employs mechanical amplitude equation and (c) the analytical transfer function model. [A.U. = Arbitrary Unit]

$$\tilde{\sigma}(\omega_T) = 2\tilde{a}^*(\omega_T)\tilde{b}(\omega_T)\tilde{c}(\omega_T) + \tilde{b}^*(\omega_T)\tilde{c}^2(\omega_T) + \tilde{a}^*(\omega_T)\tilde{c}^2(\omega_T) \quad (18a)$$

$$\sigma(t) = \text{IFT}\{\tilde{\sigma}(\omega_T)\}, \text{ for } t > T_3 \quad (18b)$$

where the three terms corresponds to group C in table 1. The quantity $\sigma(t)$ is proportional to the signal produced by the system, for $t > T_3$ under the simplifying assumption that the inhomogeneous broadening is much larger than the spectral spread of the terms in 18(a), as shown in section 2.2 [eqns 8-12]. We have also shown in section 2.2 that this term is proportional to $\tilde{\rho}_{12}(t)$, the off-diagonal density matrix element in the rotating wave basis. In fig 4(c) we have shown that implementing the transfer functions of eqn. 18 essentially yields the same result as the numerical model (for $t > T_3$, which is the relevant time span for the correlator), but at about $2 \times 10^5$ times faster simulation time for this particular case. To compare the correlation peak values for both models, let us define the ratio of the magnitude of the writing pulse, a(t) to the magnitude of the output correlation peak, s(t) as η. For the numerical model, η is 3.95 and for the analytical model, η is 3.99. The other nonlinear terms also maintain same magnitude ratio in both models.

Now, consider the situation where the reference pulse and the query pulse are each replaced by a short sequence of pulses as shown in figure 5(a). Fig 5(b) and fig 5(c) show the simulation results of the corresponding temporal correlator using the numerical model and the analytical model, respectively. Here, first we apply a very weak write pulse, $a(t) = A(t-T_1)$. Next we apply the query stream, $b(t) = B(t-T_2)$ and after some time lag the reference stream, $c(t) = C(t-T_3)$ is applied. The final correlation signal appears at time $t = -T_1 + T_2 + T_3$ inside the dotted box. It is obvious from the simulation that both models yield essentially the same result. However, the analytical model is much faster. Note that the other nonlinear terms appearing in this case do not interfere with the final correlation signal. This issue is discussed in greater detail later on in this paper.

Figure 6 shows the limiting case of the STC where the query event and the database event are each a single frame and they match exactly in the spatial domain. Specifically, the write pulse, A(x,y,t) is applied (centered) at $t = T_1$. In the time domain, it is a $\pi/2$ pulse, while spatially it is a gaussian spot (centered) at $x = 0, y = 0$. The query image, B(x,y,t) is applied (centered) at $t = T_2$, and the reference image C(x,y,t) is applied (centered) at $t = T_3$. As expected, a correlation peak appears at $t = T_2 + T_3 - T_1$. The correlation signal is computed in two different ways: first by using the explicit numerical model described in section 2.1, and then using the analytical model defined by equations 15-17, with close agreements. Here other nonlinear terms are ignored in both models, only showing the correlation signal.

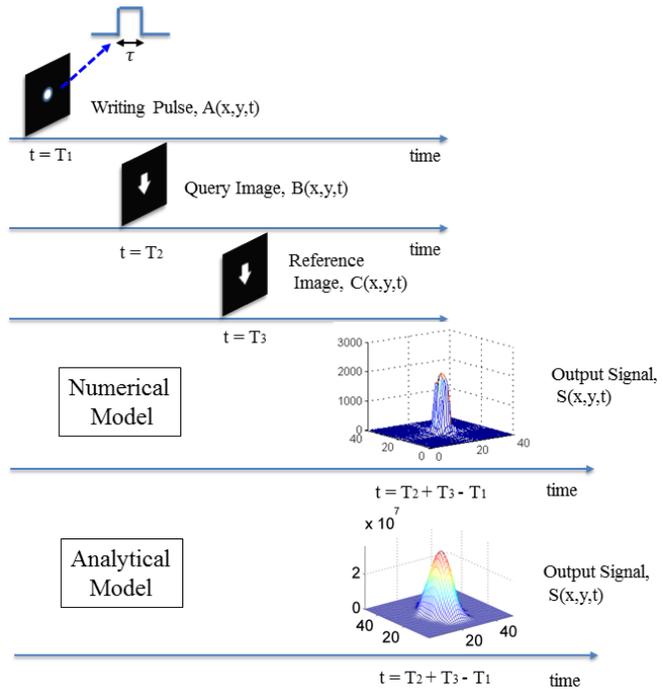

Fig 6: Simulation of the STC using the numerical model and analytical model. The output signal yields the same result for both models.

From the above simulation results, it is obvious that the numerical model and the analytical model are in good agreement with each other. Hence, we can use the analytical model reliably for simulating a three dimensional STC. Figure 7 shows the simulation result of the STC using the analytical model, where we have five query frames and ten reference frames. At time $T_1$, we apply the write pulse with a very short duration in time. In the spatial domain, the writing pulse is a small spot localized in the center of the frame. After some delay, a query video clip of five frames, denoted as B(x,y,t), are applied, at a frame rate of 30 fps (frame per second). The center of the query clips occur at time $T_2$. After additional delay, the database clips, C(x,y,t) are applied, with the same frame rate as the query frames. The database clips contain ten frames, within which only first five frames match with the query clips, as shown by the dotted box. The center of the matched clip in the reference database appears at times $T_3'$. Now, using eqns. 15-17, we get the signal S(x,y,t), which has the highest peak at time: $t = T_3' + T_2 - T_1$, as shown in the encircled feature in figure 7. Other peaks from the nonlinear terms appear at time $t = -T_2 + 2T_3$ and $t = -T_1 + 2T_3$. Now, the result can be more clearly interpreted if we integrate each frame of S(x,y,t) over space and plot it with respect to time, as shown in figure 8. Figure 8(a) shows the result for the STC considered in figure 7, where the query frames match the first half of the reference frames. It is clear from figure 8(a) that the highest peak occurs at time $t = T_3' + T_2 - T_1$. If the query frames occurred in the last half of the reference frames, the correlation peak would move 5 frames to time $t = T_3'' + T_2 - T_1$ (where $T_3' - T_3'' = (5/30)\sec$) as seen in fig 8(b).

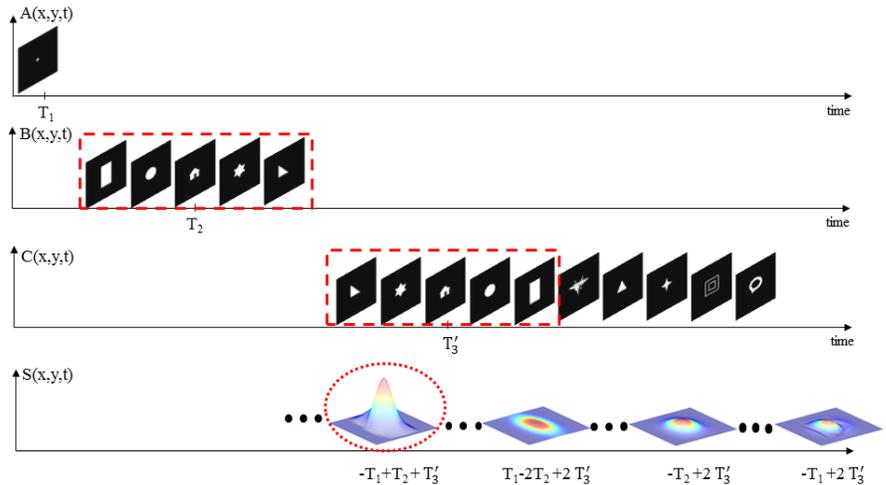

Fig 7: Simulation result of the Spatio Temporal Correlator using the analytical model. A(x,y,t) is the writing pulse, B(x,y,t) is the query frame set, and C(x,y,t) is the database frames. S(x,y,t) shows the results of the correlator at the detector plane at different time.

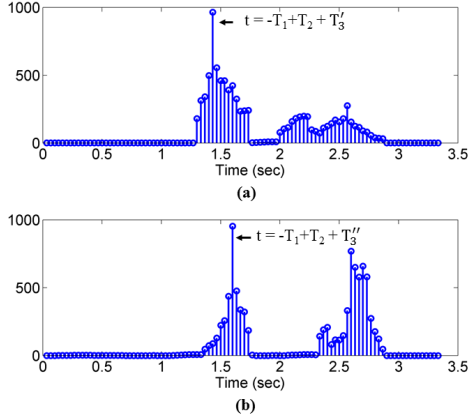

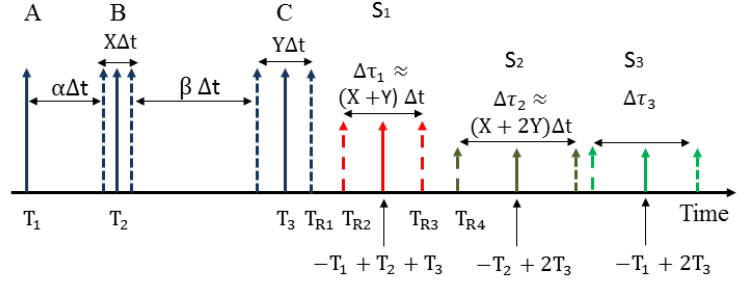

Fig 8: Simulation result of the STC where each frame integrated over space and plotted with respect to time.

Fig 9: Sequence of pulses appearing in the Spatio-Temporal Correlator.

Next, we describe how the STC can be operated in a manner such that the correlation window does not overlap the reference frames and also the nonlinear terms do not overlap the correlation signal window. The input and the output signals appearing in the STC within these constraints are illustrated schematically in figure 9. In the STC process, the center of the input pulses A(x,y,t), B(x,y,t) and C(x,y,t) appear at times $T_1$, $T_2$ and $T_3$ respectively. The write pulse, A(x,y,t) consists a single frame whereas the query, B(x,y,t) and the reference, C(x,y,t) contain multiple frames. Let us assume that, the query and the reference contain X and Y number of frames, respectively, where Y = nX (assume n ≥ 1). The peak of the correlation signal (denoted as $S_1$) appears at times $t = -T_1 + T_2 + T_3'$ where $T_3'$ is time corresponding to the center of the matching clip within the reference frames. Thus, if $T_3' = T_3$, then the correlation peak will appear at time $t_0 = -T_1 + T_2 + T_3$. However, if $T_3' < T_3 (T_3' > T_3)$ then the correlation peak will appear at a time $t < t_0 (t > t_0)$. If we assume that the matching clip is fully contained within the reference frames, then it follows that the location of the correlation peak will be confined to a window of width $\sim (Y-X)\Delta t$, where $\Delta t$ is the sum of the duration of each frame and the gap between the adjacent frames. However, convolution between two matching signals (i.e. the query and the matching clips), each with a duration of $\sim X\Delta t$, produces a signal with a temporal width of $\sim 2X\Delta t$. Thus, the overall window for observing the correlation signal will have a width $\sim (X+Y)\Delta t \equiv \Delta \tau_1$, centered at time $t_0 = -T_1 + T_2 + T_3$.

Let us denote as $\alpha \Delta t$ the time lag between the writing pulse, A(x,y,t), and the first frame in the query clip, B(x,y,t). The value of $\alpha$ has to be chosen to ensure that the observation window for the correlation signal does not overlap the reference frames, C(x,y,t). Let us assume that, the rightmost frame of the reference, C(x,y,t) occurs at time $T_{R1}$ and the leftmost edge of the correlation window occurs at time $T_{R2}$. We thus need to ensure that $T_{R2} - T_{R1} \geq 0$. Considering $T_1 = 0$ for simplicity and without loss of generality and noting that the width of the correlation window is $\Delta \tau_1 \approx (X+Y)\Delta t$, it is easy to show that $\alpha \geq nX$ satisfies this constraint. For example, if the number of query frames is X = 3 and the number of reference frames is Y = 9, then n = 3 and $\alpha$ should be chosen greater than or equal to 9.

As mentioned earlier, we have investigated the contribution of additional non-linear terms in section 2.2. Specifically, in Table 1 of section 2.2, we have cataloged all the non-linear terms. Given that we look for signal only after applying the reference frames, the only terms of relevance are those shown in group C of this table. There are three terms here. One of these is the desired correlation signal, denoted as $S_1 = IFT(\tilde{a}^*\tilde{b}\tilde{c})$ where $\tilde{a}, \tilde{b}, \tilde{c}$ are the three dimensional FT's of A(x,y,t), B(x,y,t) and C(x,y,t) respectively. The other two are: $S_2 = IFT(\tilde{b}^*\tilde{c}^2)$, $S_3 = IFT(\tilde{a}^*\tilde{c}^2)$ which will appear centered at $t = 2T_3 - T_2$ and $t = 2T_3 - T_1$, respectively. Using the same set of arguments we used earlier to determine the width of the observation window for the correlation signal, we can show that the width of the window for $S_2$ is $\sim (X+2Y)\Delta t \equiv \Delta\tau_2$ and the width of the window of window for $S_3$ is $\sim (Y+1)\Delta t \equiv \Delta\tau_3$. Thus, if we ensure that the window for $S_2$ does not overlap that of $S_1$, that will also ensure that there is no overlap between the windows for $S_3$ and that of $S_1$.

Let us denote $\beta\Delta t$ as the temporal separation between the rightmost frame of the query clip and the leftmost frame of the reference. We denote $T_{R3}$ as the right edge of the window for $S_1$, and $T_{R4}$ as the left edge of the window for $S_2$. We need to choose $\beta$ to have a value such that $T_{R4} \geq T_{R3}$. It is easy to show that this condition is satisfied if $\beta \geq (2nX + X)$. For example, if the number of query frames is X=3 and the number of reference frames is Y = 9, then n = 3 and $\beta$ should be chosen to be greater than or equal to 21 to avoid overlap between the windows for $S_1$ and $S_2$.

## 6. Practical Considerations for Realizing an Automatic Target Recognition System:

### 6.1 Counter Propagating Scheme using a Three-level $\Lambda$ system:

In the analysis presented above, we have considered a two level system without decay. Here, we describe how such a system can be realized effectively by making use of a three level $\Lambda$ system, as shown in figure 10.

A particular example of such a system, based on $^{87}$Rb atoms, consists of levels $|1\rangle$ ($5^2S_{1/2}$; F=1), $|2\rangle$ ($5^2S_{1/2}$; F=2), and $|3\rangle$ ($5^2P_{1/2}$ manifold). If the two optical fields are highly detuned, this system behave effectively as a two

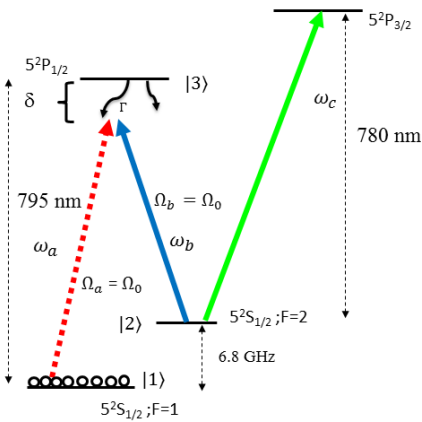

Fig 10: Raman interaction in a three level system

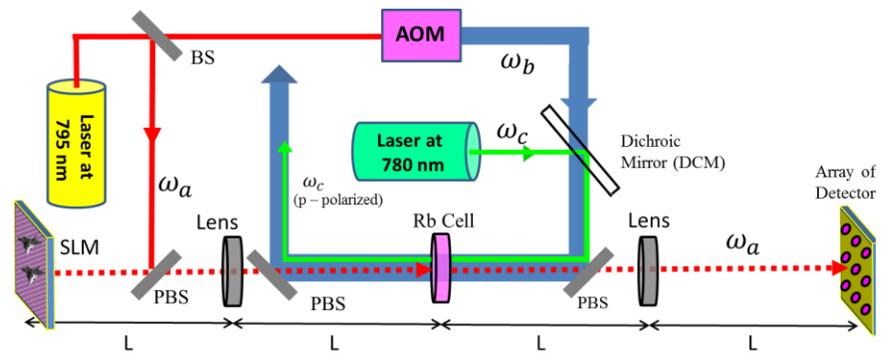

Fig 11: Schematic illustration of the physical implementation of the Automatic Event Recognition system employing vapor cell containing $^{87}$Rb atoms and neon buffer gas. [PBS = Polarizing Beam Splitter; AOM: Acousto-Optic Modulator; SLM: Spatial Light Modulator]. See text for details.

level system consisting of states $|1\rangle$ and $|2\rangle$[21]. The effective 1-2 transition is inhomogeneously broadened due to Doppler shifts. If the two optical beams are co-propagating, this Doppler width at room temperature is very small. However, if the beams are counter-propagating, then this Doppler width is quite large. As such, we must make use of the counter-propagating scheme if we want to maximize the bandwidth of the system. The use of this $\Lambda$ system for realizing the AER requires significant modifications of the conceptual architecture shown earlier in figure 3. In figure 11, we show the physical implementation of the AER architecture employing the $\Lambda$ transition in $^{87}$Rb atoms in a vapor cell. A pulsed auxiliary beam (at frequency $\omega_c$) that couples level $|2\rangle$ to the $5^2P_{3/2}$ manifold is used to optically pump the atoms into level $|1\rangle$ before any correlation process starts.

As noted above, this geometry corresponds to maximization of the inhomogeneous broadening of the effective two level transition between states $|1\rangle$ and $|2\rangle$, which we denote here as $\Gamma_{INH}$. In general, for high fidelity operation of the system, it is necessary to ensure that this broadening is greater than the inverse of the duration of the shortest pulse in the data stream, which we denote here as $\Gamma_P$. On the other hand, if $\Gamma_{INH} \gg \Gamma_P$, then only a small fraction (of the order of $\Gamma_P/\Gamma_{INH}$) contributes effectively to the correlation signal. Thus, under such a scenario, it is useful to employ a scheme where the angle of the beam at frequency $\omega_b$ (with respect to the direction of propagation of the beams at frequency $\omega_a$) is varied to make the value of $\Gamma_P/\Gamma_{INH}$ close to unity. Specifically, let us denote this angle as $\theta_{ab}$, as shown in figure 12 (Note that figure 11 can be interpreted to be the same as figure 12, with the value of $\theta_{ab}$ being $180^0$). In general, the value of $\Gamma_{INH}$ as a function of $\theta_{ab}$ is given by $\Gamma_{INH} = 2[\omega_a - \omega_b Cos(\theta_{ab})]u/c$, where $u$ is the most probable velocity in the vapor cell, and $c$ is the speed of light. Assuming a typical value of u=250 m/sec, and ignoring the small difference between the values of $\omega_a$ and $\omega_b$, we get that $F_{INH} \equiv \Gamma_{INH}/2\pi \approx [1 - Cos(\theta_{ab})]*630 MHz$. Thus, for the configuration shown in figure 11, where $\theta_{ab}$ is 180 degrees, the inhomogeneous width is ~1260 MHz. If we use the nearly co-propagating geometry, with $\theta_{ab}$ equaling 15 degrees for example, the inhomogeneous width would be ~21.5 MHz. For the case of exact co-propagation, we must take into account the small difference between $\omega_a$ and $\omega_b$, and in that case we have $F_{INH} \approx 11\, kHz$. In practice, the choice of $\theta_{ab}$ would be dictated by the need to ensure that $\Gamma_P/\Gamma_{INH}$ is close to unity in order to maximize the efficiency of the correlation process.

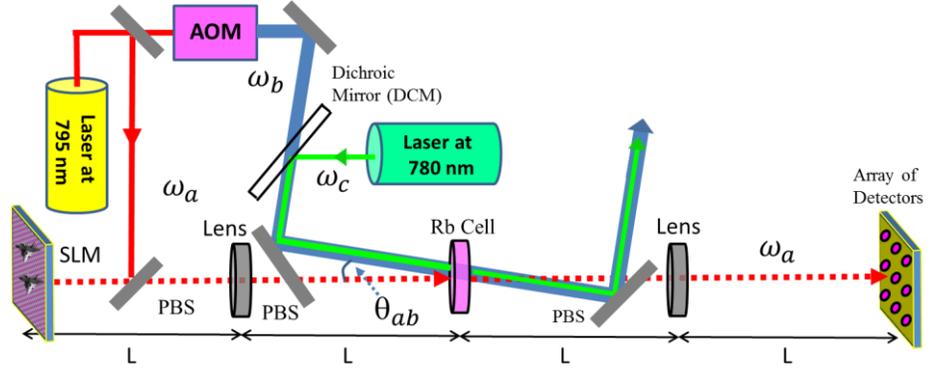

Figure 12: Schematic illustration of the physical implementation of the Automatic Event Recognition system employing paraffin coated porous glass vapor cell containing [87]Rb atoms, with nearly co-propagating Raman interaction. [PBS = Polarizing Beam Splitter; AOM: Acousto-Optic Modulator; SLM: Spatial Light Modulator]. See text for details.

The beams (at frequency $\omega_a$) that carry the image information as well as the recording pulse are applied along the 1-3 transition, detuned by an amount, $\delta$. The beam (at frequency $\omega_b$) that excites the 2-3 transition is applied at all times, is also detuned by the same amount, $\delta$. The value of $\delta$ is chosen to be much larger than the decay time of level $|3\rangle$, in order to ensure that the three-level system function effectively as a two level system coupling state $|1\rangle$ to state $|2\rangle$ [21]. The two Raman transition beams are polarized to be linear and orthogonal to each other. The optical pumping beam at frequency $\omega_c$ has the same linear polarization as that of the Raman beam at frequency $\omega_b$. A dichroic mirror is used to combine these two frequencies, made possible by the fact that they differ in wavelengths by ~15 nm. The correlation signal appears at frequency $\omega_a$, and passes through the second Polarizing Beam Splitter (PBS). Finally, the correlation signal is observed by the detector array. The resulting voltage signals from all the detectors are integrated to produce the net signal of the AER system. Alternatively, a single detector with a large area to replace the array can be used to produce the net signal automatically.

### 6.2    Memory time of the atomic media:

A key feature of the atomic medium is that it stores the spatial and temporal interference between the recording pulse and the query frames in the electro-nuclear spin coherence in the form of a coherent superposition between states $|1\rangle$ to $|2\rangle$. The lifetime of this coherence time can be ~ 1 second in a paraffin coated Rb vapor cell[30,31]. The spatio-temporal correlation process must be carried out within this time window. However, as we discuss later, this time window does not limit the maximum size of the data base video that one can search through.

Consider a situation where the query clip has a nominal duration ($T_{NOM}$) of 20 seconds, if played on a regular monitor. For a video frame rate ($F_{VFR}$) of 30 per second, this would contain 600 frames. For the SLM loading speed of 600 microseconds (defined as $T_{SLM}$) per frame, this clip can be loaded into the atomic medium in 360 ms, which is much shorter than the nominal duration (20 seconds) of the query clip. Thus, in the context of the AER, the retrieval duration of a video clip is given by $T_{RET} = F_{VFR} * T_{NOM} * T_{SLM}$.

As we mentioned above, it is very important to note that the atomic coherence memory time of 1 second does not constrain the size of the database video one can search through. To see this, consider a situation where the memory time (i.e. the coherence time) is $T_3$, the time span of the query clip (expressed as its retrieval duration) is $T_1$, and the time span for the database video (expressed as its retrieval duration) is $T_2$. The correlator is operated for the duration $T_3$, during which a fraction (given by $T_3/T_2$) of the database has been searched. At this point, the AER system is reinitialized by using the optical pumping beam at frequency $\omega_c$ (see figures 10, 11 and 12), and the same query clip is loaded again. The database is now loaded with a start time of ($T_3-T_1$), the AER is operated for another duration of $T_3$, and the process is repeated again, with a start time of $2T_3-T_1$, and so on, until the whole data base has been searched.

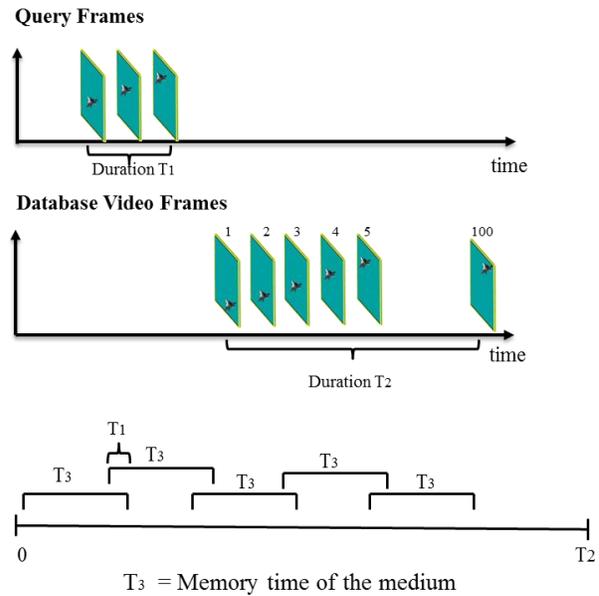

Figure 13: Schematic illustration of the process of the time sequencing necessary for the AER system. See text for details.

This sequencing is illustrated schematically in figure 13. The offset of $T_1$ in the start time of the database is to ensure that the AER would be able to detect the presence of the query clip even if it occurs in-between each segment searched within each memory window.

As noted earlier, the smallest loading time that can be accommodated by the atomic memory is determined by the effective inhomogeneous width ($F_{INH}$) of the two-photon transition coupling level $|1\rangle$ to level $|2\rangle$. When the two beams that cause this transition (at $\omega_a$ and $\omega_b$) are counter-propagating, as shown in figure 11, the inhomogeneous width becomes ~1.26 GHz, and this can accommodate a loading time as fast as 1 ns.

### 6.3 Implementing AER employing a vapor cell, a holographic video disc, and a lithium niobate crystal:

One possible way to make use of such a short loading time is to employ a very fast SLM. In recent years, significant progress has been made in developing technologies that could lead to the development of high speed SLMs. One example of such a development is a quantum well structure[32,33] that can be modulated very fast (>50 GHz) and efficiently to produce phase modulation. However, in order to achieve this frame loading rate in practice, it is necessary to create a computer architecture and a data bus that can retrieve a page of data in a microsecond and transfer it to the SLM. While in principle it is possible to build such a system, it does not currently exist, to the best of our knowledge. Thus, in order to make use of the fast loading time allowed by the atomic medium, one can explore the use of holographic techniques. The basic approach to be employed for this makes use of a holographic video disc, which is discussed in detail in reference 22. Since the query clips have to be periodically refreshed, based on the event of interest, one can make use of a rewritable memory, namely a lithium niobate crystal[34,35].

During the operation of the AER system, the data read from the crystal will be made co-linear with the data read from the holographic video disc using a 50/50 beam splitter, as illustrated in figure 14. To access different frames within the query clip, an acousto-optic deflector (AOD) can be used to scan the read angle. Typical speed of operation for such a device is about 20 MHz, with the number of resolvable angles being as high as 5000. Thus, it should be possible to retrieve images from the query clip at the rate of 1 frame per microsecond (which is much less than the maximum possible rate of 20 frames per microsecond corresponding to a bandwidth of 20 MHz). To access different frames from the holographic video disc (i.e., the video data base), at a given location on the disc, again one can use an acousto-optic deflector, with the same retrieval rate (i.e., about 1 frame per microsecond). To access different locations sequentially, a combination of a motorized rotator and a motorized translator can be used. For a 1 cm thick disc, the Bragg angular selectivity can be about 0.002 degrees. The angular range covered by a well-designed AOD can be ~10 degrees, so that about 5000 images can be read-out from one location. This is much larger than the 900 images in a query clip of 30 seconds of physical duration. Thus, using the sequencing process illustrated in figure 13, the AER would be operated for a duration that is long enough to load the 5000 images from one location, corresponding to ~ 5 msec (i.e., $T_3$=5 msec in the notation used in figure 13). The atomic medium will then be re-initialized by using the optical pumping beam at frequency $\omega_c$, while the rotation stage is used to move the system to the next location on the disc. The AER operation will then be repeated with the data retrieved from the new location. Once all locations at the same radial distance from the center of the disc have been searched through, the translation stage can be used to enable the AOD to access data from the locations at a different (adjacent) radial distance. If we assume conservatively that moving from one location to another can be accomplished in a stable manner in about 30 msecs, then all 500 locations can be searched in less than 20 seconds. In order to realize this approach it is necessary to make holographic video discs with a thickness of 1 cm. One material suitable for realizing such a disc is made of PMMA (polymethyl methalycrate), doped with the dye called phenanthren-quinone[36,37,38,39,40,41,42,43,44].

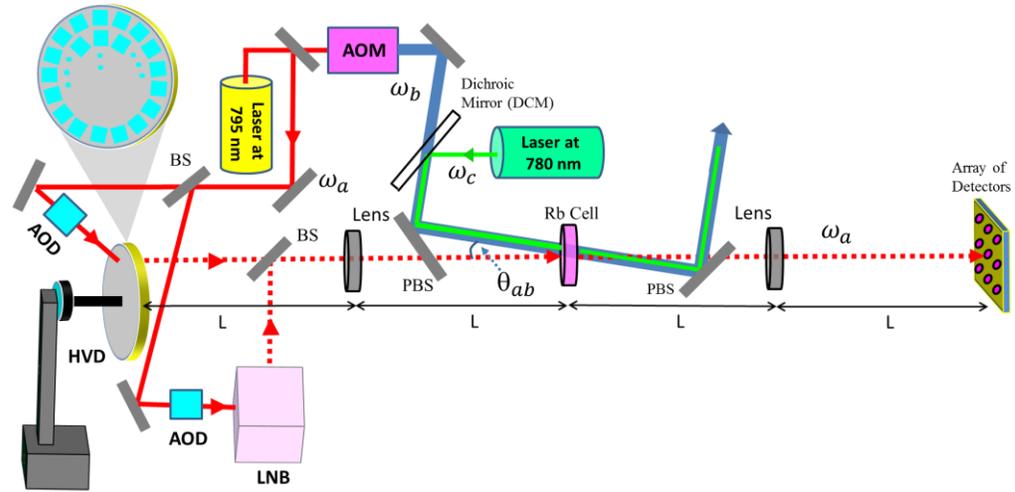

Figure 14: Schematic illustration of the physical implementation of the Automatic Event Recognition system employing paraffin coated porous glass vapor cell containing 87Rb atoms, with nearly co-propagating Raman interaction, and using a holographic video disc and a lithium niobate crystal for rapid loading of data base and query videos. [PBS = Polarizing Beam Splitter; AOM: Acousto-Optic Modulator; AOD: Acousto-Optic Deflector; HVD: Holographic Video Disc; LNB: Lithium Niobate Crystal]. See text for details.

For a typical imaging system, the spatial resolution is about 10 um X 10 um. Thus, it is important to ensure that the atoms stay localized to such an area during each stage of the correlation sequence shown in figure 13. In order to meet this constraint, one can make use of a vapor cell consisting of nano-porous glass[45,46,47]. Specifically, the process will make use of porous glass (PG) which is about 96% silica, and contains randomly inter-connected pores with a mean diameter that can be controlled to be anywhere from 5 nm to 40 nm. A typical PG plate available from Schott, Inc., has a dimension of the order of 2 cm X 2 cm X 1 mm. Using a simple variation of the technique described in ref. 31, the inner walls of the pores in the PG can be coated with paraffin in order to ensure that the Rb vapor does not lose coherence when they collide with the walls of the pores[30,31].

### 6.4 Detection of the correlation signal:

For detection of the correlation signal, one can make use of an image intensified high CCD camera, which has a speed of 200 psec. The signal captured by the camera can be acquired rapidly using a PCIe bus. As noted earlier, the signature of a positive match for a clip is given by the signal integrated over the detector array. As such, the signals captured from the camera can be integrated by the computer, and displayed on the screen. Alternatively, to eliminate the speed bottleneck imposed by this integration process, one can use a single detector with a large area to produce the signal.

### 7. Conclusion

We describe an automatic event recognition (AER) system based on a three-dimensional spatio-temporal correlator (STC) that combines the techniques of holographic correlation and photon echo based temporal pattern recognition. The STC is shift invariant in space and time. It can be used to recognize rapidly an event (e.g., a short video clip) that may be present in a large video file, and determine the temporal location of the event. Using polar Mellin transform, it is possible to realize an STC that is also scale and rotation invariant spatially. Numerical simulation results of such a system are presented using quantum mechanical equations of evolution. For this simulation we have used the model of an idealized, decay-free two level system of atoms with an inhomogeneous broadening that is larger than the inverse of the temporal resolution of the data stream. We show how such a system can be realized by using a lambda-type three level system in atomic vapor, via adiabatic elimination of the intermediate state. We have also developed analytically a three dimensional transfer function of the system, and shown that it agrees closely with the results obtained via explicit simulation of the atomic response. The analytical transfer function can be used to determine the response of an STC very rapidly. In addition to the correlation signal, other nonlinear terms appear in the explicit numerical model. These terms are also verified by the analytical model. We develop a way to detect the correlation signal without interference from these additional nonlinear terms. We also show how such a practical STC can be realized using a combination of a porous-glass based Rb vapor cell, a holographic video disc, and a lithium niobate crystal.

### 8. Acknowledgements

This work is supported by AFOSR Grant FA9550-10-01-0288.

## Appendix A: Transfer Function for the Spatio-Temporal Correlator

Consider the complete spatio-temporal correlator system, shown in figure A.1. Here, the signals are generated by modulating the field from a laser with a spatial light modulator (SLM), for example. Thus, each of three pulses are encoded with two dimensional spatial information. Specifically, we now have three functions containing information: $A(x,y,t)$, $B(x,y,t)$ and $C(x,y,t)$, where $(x,y)$ are the rectilinear spatial coordinates in the plane of the SLM. The corresponding signals in the plane of the atomic medium are Fourier Transformed (FT'd) in the spatial domain due to the first lens. Before writing these functions down, we recall briefly the mathematical relations between a spatial function, $U(x,y)$, in the SLM plane, and the corresponding spatial function $V(x_a, y_a)$ in the atomic medium plane, which has coordinates $(x_a, y_a)$. We consider first the two dimensional FT of the function $U(x,y)$, denoted as $\tilde{U}(k_x, k_y)$, which are related to each other as follows:

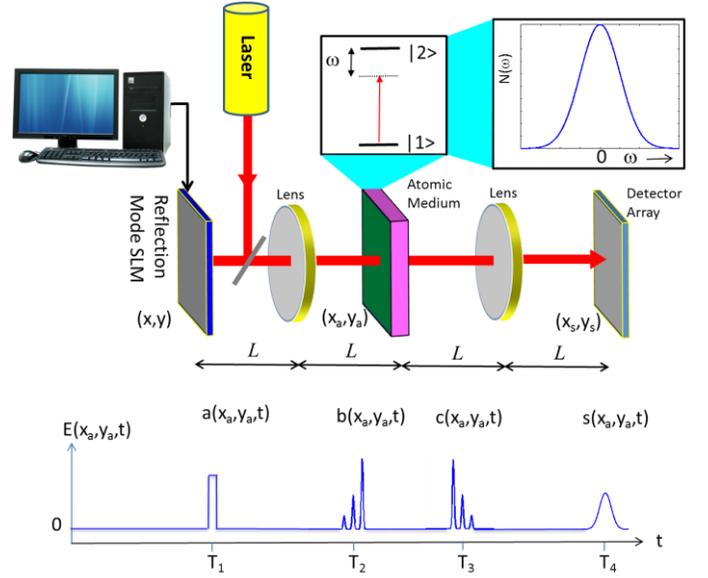

Fig A.1: Illustration of the spatio-temporal correlator. Here, the focal length of each lens is L. See text for details.

$$\tilde{U}(k_x, k_y) = \frac{1}{2\pi}\int_{-\infty}^{\infty} dx \int_{-\infty}^{\infty} dy\, U(x,y)\exp[-i(k_x x + k_y y)]; \quad U(x,y) = \frac{1}{2\pi}\int_{-\infty}^{\infty} dk_x \int_{-\infty}^{\infty} dk_y\, \tilde{U}(k_x, k_y)\exp[i(k_x x + k_y y)] \quad \text{(A1)}$$

According to the laws of Fresnel diffraction, and properties of ideal lenses, the function $V(x_a, y_a)$ is given by:

$$V(x_a, y_a) = \frac{\exp[i2kL]}{i\lambda L}\tilde{U}(k_x, k_y)\bigg|_{k_x = \frac{2\pi x_a}{\lambda L}, k_y = \frac{2\pi y_a}{\lambda L}} = \frac{\exp[i2kL]}{i\lambda L}\tilde{U}\left(\frac{2\pi x_a}{\lambda L}, \frac{2\pi y_a}{\lambda L}\right) \quad \text{(A2)}$$

where L is the focal length of the lens, and λ is the wavelength of the laser. Thus, aside from the inconsequential phase factor and the scaling factor, the spatial function in the plane of the atomic medium is the spatial FT of the spatial function in the plane of the SLM. Similarly, the spatial function of the field produced in the plane of the detector, $W(x_s, y_s)$ which has coordinates $(x_s, y_s)$, will be the spatial FT of the spatial function in the plane of the atomic medium. Thus, in the absence of any interaction with the atomic medium, the spatial function of the field in the plane of the detector becomes an exact reproduction of the spatial function of the field in the plane of the SLM, but inverted in both x and y directions, along with a phase shift factor of $\exp(i4kL)$:

$$W(x_s, y_s)\big|_{\text{no atomic medium}} = -\exp(i4kL)U(-x,-y) \quad \text{(A3)}$$

This is the well-known 4F imaging process. It should be noted that the prefactor of $(1/i\lambda L)$ in eqn. A2 gets compensated for during the second stage, and does not appear in eqn. A3.

We now define the three-dimensional (spatio-temporal) FT of a function $g(x,y,t)$ as follows:

$$\tilde{g}(k_x, k_y, \omega) = 1/[2\pi]^{3/2} \int_{-\infty}^{\infty} dx \int_{-\infty}^{\infty} dy \int_{-\infty}^{\infty} dt\, g(x,y,t)\exp[-i(k_x x + k_y y + \omega t)] \quad \text{(A4)}$$

$$g(x,y,t) = 1/[2\pi]^{3/2} \int_{-\infty}^{\infty} dk_x \int_{-\infty}^{\infty} dk_y \int_{-\infty}^{\infty} d\omega\, \tilde{g}(k_x, k_y, \omega)\exp[i(k_x x + k_y y + \omega t)] \quad \text{(A5)}$$

Noting that the spatial FTs of the signals A(x,y,t), B(x,y,t) and C(x,y,t) appear at the plane of the atomic medium at times $T_1$, $T_2$, and $T_3$, respectively, the corresponding Rabi frequencies in the three dimensional spectral domain can be expressed as (cf. eqn 6 ):

$$\tilde{\Omega}_a(k_x,k_y,\omega) = \zeta\mu\tilde{A}(k_x,k_y,\omega)\exp(i\omega T_1)\,;\,\tilde{\Omega}_b(k_x,k_y,\omega) = \zeta\mu\tilde{B}(k_x,k_y,\omega)\exp(i\omega T_2)\,;$$
$$\tilde{\Omega}_c(k_x,k_y,\omega) = \zeta\mu\tilde{C}(k_x,k_y,\omega)\exp(i\omega T_3) \quad (A6)$$

where $\zeta = \exp(i2kL)/(i\lambda L)$ (see the prefactor in eqn. A3), and it is to be understood that the spatial frequency components $\{k_x,k_y\}$ corresponds physically to spatial locations $\{(k_x\lambda L/2\pi),(k_y\lambda L/2\pi)\}$ in the plane of the atomic medium. Using the same line of arguments and making use of the same set of approximations as presented in Section 2.2, we then conclude that the normalized signal in the plane of the detector array, corresponding to the correlation signal, is given by (cf. eqn. 13):

$$\sigma(x_s,y_s,t) = \frac{1}{[2\pi]^{3/2}}\int_{-\infty}^{\infty}dk_x\int_{-\infty}^{\infty}dk_y\int_{-\infty}^{\infty}d\omega[\tilde{A}^*(k_x,k_y,\omega)\times\tilde{B}(k_x,k_y,\omega)\tilde{C}(k_x,k_y,\omega)\exp\{i\omega(T_3+T_2-T_1)\}]$$
$$\times\exp[-i(k_x x_s + k_y y_s + \omega t)] \quad (A7)$$

It then follows that the three-dimensional FT of the normalized correlation signal is:

$$\tilde{\sigma}(k_x,k_y,\omega) = \tilde{A}^*(\omega)\tilde{B}(\omega)\tilde{C}(\omega)\exp(j\omega(T_3+T_2-T_1)) \quad (A8)$$

If we define $\tilde{S}(k_x,k_y,\omega) = \tilde{A}^*(k_x,k_y,\omega)\tilde{B}(k_x,k_y,\omega)\tilde{C}(k_x,k_y,\omega)$ then it follows that $S(x_s,y_s,t)$ is the three dimensional cross-correlation between $A(x,y,t)$ and the three-dimensional convolution of $B(x,y,t)$ and $C(x,y,t)$. Since $A(x,y,t)$ is essentially a delta function in both temporal and spatial domains (i.e., it is a very short temporal pulse, and is a small point signal at the center of the SLM plane), $S(x_s,y_s,t)$ is effectively the three-dimensional convolution of $B(x,y,t)$ and $C(x,y,t)$. Explicitly, if we consider $A(x,y,t) = A_0\delta(x)\delta(y)\delta(t)$, we get: $S(x_s,y_s,t) = A_o\int_{-\infty}^{\infty}dt'\int_{-\infty}^{\infty}dx'\int_{-\infty}^{\infty}dy'B(x',y',t')\times C(x_s-x',y_s-y',t-t')$.

If we take into account the finite width of $A(x,y,t)$ in all three dimensions, this signal $S(x_s,y_s,t)$ will be broadened by these added widths in each dimension. Finally, we note that $\sigma(x_s,y_s,t) = S(x_s,y_s,t-(T_3+T_2-T_1))$, which means that this correlation signal occurs at $t=T_3+(T_2-T_1)$, in the plane of the detector array.